\documentclass[fleqn,usenatbib]{mnras}

\usepackage[T1]{fontenc}
\usepackage{ae,aecompl}

\usepackage{graphicx}	
\usepackage{amsmath}	
\usepackage{amssymb}	

\usepackage{mathrsfs,dsfont}
\usepackage{multirow}
\usepackage{captcont,subcaption}
\usepackage{float}
\usepackage{booktabs}
\usepackage{epstopdf}
\usepackage{color}
\usepackage{hyperref}
\usepackage{verbatim}

\usepackage[none]{hyphenat}

\graphicspath{{figures/}}

\def\be{\begin{equation}}
\def\ee{\end{equation}}

\def\Myr{{\rm \,Myr}}
\def\Gyr{{\rm \,Gyr}}

\def\kpc{{\rm \,kpc}}

\def\keV{{\rm \,keV}}

\def\msun{{\,M_\odot}}

\newcommand{\dd}{{\rm d}}

\title[Runaway merger shocks in galaxy clusters]{Runaway Merger Shocks in Galaxy Cluster Outskirts and Radio Relics}

\author[Congyao Zhang et al.]{
Congyao Zhang,$^1$\thanks{E-mail: cyzhang@mpa-garching.mpg.de}
Eugene Churazov,$^{1,2}$
William R. Forman,$^{3}$
Natalia Lyskova$^{4,2}$
\\
$^1$~Max Planck Institute for Astrophysics, Karl-Schwarzschild-Str. 1, D-85741 Garching, Germany  \\
$^2$~Space Research Institute (IKI), Profsoyuznaya 84/32, Moscow 117997, Russia \\
$^3$~Smithsonian Astrophysical Observatory, Harvard-Smithsonian Center for Astrophysics, 60 Garden St., Cambridge, MA 02138 \\
$^4$~National Research University Higher School of Economics, Myasnitskaya str. 20, Moscow 101000, Russia
}

\date{Accepted XXX. Received YYY; in original form ZZZ}

\pubyear{2019}

\begin{document}
\label{firstpage}
\pagerange{\pageref{firstpage}--\pageref{lastpage}}
\maketitle

\begin{abstract}
Moderately strong shocks arise naturally when two subclusters merge. For instance, when a smaller subcluster falls into the gravitational potential of a more massive cluster, a bow shock is formed and moves together with the subcluster. After pericenter passage, however, the subcluster is decelerated by the gravity of the main cluster, while the shock continues moving away from the cluster center. These shocks are considered as promising candidates for powering radio relics found in many clusters. The aim of this paper is to explore the fate of such shocks when they travel to the cluster outskirts, far from the place where the shocks were initiated. In a uniform medium, such a ``runaway'' shock should weaken with distance. However, as shocks move to large radii in galaxy clusters, the shock is moving down a steep density gradient that helps the shock to maintain its strength over a large distance. Observations and numerical simulations show that, beyond $R_{500}$, gas density profiles are as steep as, or steeper than, $\sim r^{-3}$, suggesting that there exists a ``Habitable zone'' for moderately strong shocks in cluster outskirts where the shock strength can be maintained or even amplified. A characteristic feature of runaway shocks is that the strong compression, relative to the initial state, is confined to a narrow region just behind the shock. Therefore, if such a shock runs over a region with a pre-existing population of relativistic particles, then the boost in radio emissivity, due to pure adiabatic compression, will also be confined to a narrow radial shell.
\end{abstract}




\section{Introduction} \label{sec:introduction}

Shocks naturally form during the merger process of galaxy groups and clusters and frequently lead to remarkable observational signatures, e.g. sharp X-ray surface brightness edges, gas temperature discontinuities, and radio relics (see \citealt{Markevitch2007,Feretti2012,Bykov2015,Weeren2019} for reviews). In X-ray observations, these shocks are usually found well within the cluster virial radius, and have a moderate Mach number ($\mathcal{M}_{\rm s}\sim{\rm few}$), in contrast to the cluster virial shocks ($\mathcal{M}_{\rm s}\sim10-10^3$; see e.g. \citealt{Borgani2011}).

\begin{figure*}
\centering
\includegraphics[width=0.9\linewidth]{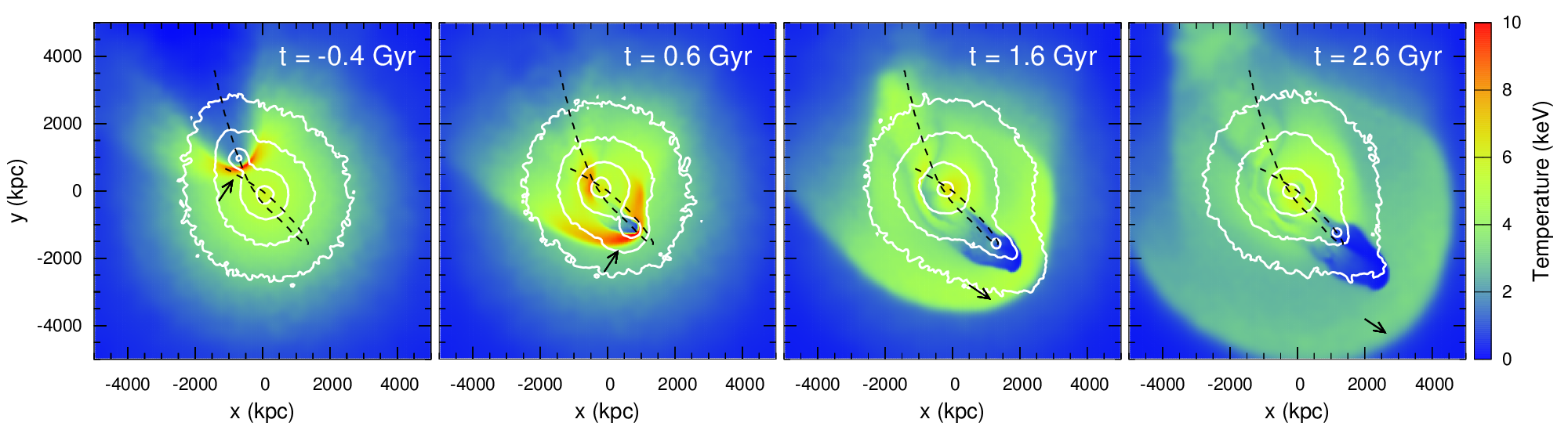}
\caption{Simulated evolution of the X-ray-weighted temperature distribution during a merger of two idealized subclusters. This simulation was performed to illustrate the merger of the galaxy group around NGC~4839 in the Coma cluster and the formation of an associated runaway shock as discussed by \citet{Lyskova2019}. A small subcluster enters into the main cluster from the top-left (the mass ratio between these two merging objects is $10$; see \citealt{Lyskova2019} and also \citealt{Zhang2014} for more information on the simulation). We set the evolution time $t=0$ at the primary pericentric passage. The black dashed line shows the trajectory of the infalling small subcluster; the white contours show the mass surface density of the merging system. The black arrows mark the position of the bow shocks. This figure illustrates the formation of a runaway merger shock during the cluster merging process: in the first two panels, the merger shock is in the ``driven phase'' -- it moves together with the subcluster; in the third and fourth panels, the bow shock detaches from the subcluster and moves towards the main-cluster outskirts (see Section~\ref{sec:introduction}). In our idealized simulations, the outer regions of the main cluster are initially very close to hydrostatic equilibrium. This leads to a very smooth and well-behaved shape of the shock at late stages of its evolution (see e.g. the right-most panel at $t=2.6\Gyr$). In cosmological simulations, cluster outskirts are much more perturbed, leading to irregularities or even complete destruction of shocks, especially in the regions where dense filaments are infalling to the cluster. Nevertheless, the outgoing shock can maintain its shape in the ``valleys'' between the filaments (see simulation results e.g. \citealt{Paul2011}). At smaller distances from the cluster core (between $R_{500}$ and $R_{\rm vir}$), there are observational examples of Mpc-long shocks, e.g. in the ``Sausage'' cluster (see e.g. \citealt{Weeren2010}). }
\label{fig:coma_simulation}
\end{figure*}

These merger shocks have also been extensively explored through numerical methods (e.g. \citealt{Ricker2001,Vazza2009,Paul2011,Ha2018}). In general, the evolution of merger shocks in galaxy clusters could be summarized as a two-phase process (see Fig.~\ref{fig:coma_simulation} for an example):
\begin{itemize}
  \item In the initial ``driven phase'', a bow shock is formed ahead of the gaseous core of the infalling subcluster\footnote{The boundary separating the gaseous atmosphere of the main cluster from that of the subcluster is a contact discontinuity, a.k.a. cold front.}.  \citet{Zhang2019} have made a systematic study of this phase, and have shown that the velocities of the cold front, shock, and dark matter (DM) halo of the subcluster could be quite different, especially after the passage of the subcluster through the main cluster core. This is caused by (1) the acceleration/deceleration of the subcluster while moving in the gravitational potential well of the main cluster; and (2) the density/pressure gradients of the intracluster medium (ICM) of the main cluster.
  \item In the second phase, the shock velocity deviates strongly from the velocity of the subcluster (in amplitude and/or direction), for example, the velocities differ when the subcluster is decelerated by the gravity of the main cluster after core passage, and finally falls backwards after crossing the apocenter (see the right panel in Fig.~\ref{fig:coma_simulation}). The merger shock consequently transits into a ``detached phase'' -- the bow shock detaches from the body which drives it (i.e. the gas core of the infalling subcluster) and moves farther towards the main-cluster outskirts. In this phase, the subcluster has only limited impact on the runaway shock properties, whose propagation is mainly determined by the shock Mach number and the density/temperature profiles of the main cluster.
\end{itemize}

In this paper, we concentrate on the fate of merger shocks in the ``detached phase''. Unlike the case of a shock moving in a homogeneous medium, the steep gas density gradient of the main cluster atmosphere helps the runaway shock maintain its strength over a long distance (even possibly be amplified, if the gradient is sufficiently steep).

The current/recent generation of X-ray observatories (e.g. \textit{Chandra}, \textit{XMM-Newton}, Suzaku) have routinely measured the gas density profiles of galaxy clusters within $\sim R_{500}$ (the radius within which the mass density is 500 times the critical density at the cluster redshift) and sometimes beyond (see e.g. \citealt{Vikhlinin2006,Bautz2009,George2009,Morandi2015,Ghirardini2019,Okabe2014}).
Observations of the Sunyaev-Zel'dovich (SZ) effect (e.g., \textit{Planck}, \textit{SPT}) extend the radial range to a few virial radii ($R_{\rm vir}\simeq 1.7R_{500}$). In general, the gas density profiles of galaxy clusters are fairly flat in the central region ($\sim r^{-1}$), and become progressively steeper with radius (e.g. $\sim r^{-2.5}$ around $R_{500}$).
This implies that, beyond $\sim R_{500}$, a ``Habitable zone'' of runaway merger shocks exists, where shocks are able to maintain their strength and are, therefore, ``long-lived''.

The aim of this work is to understand the behaviors of the runaway merger shocks in galaxy clusters (particularly in the ``Habitable zones''). The radio relics discovered in cluster peripheries are possibly associated with these ``long-lived'' runaway shocks \citep[see e.g.][]{Vazza2012,Hong2014,Kang2015,Ha2018}, which, on the other hand, provide a unique opportunity to identify and study these shocks whose X-ray signature has become too weak to be detected. In this study, instead of full cosmological runs, we use idealized/controlled simulations that elucidate conditions required for ``survival'' of the shocks. A one-dimensional (1D) blast wave, propagating through a power-law density distribution, is the simplest model that captures the essential properties of runaway shocks and probes a wide range of shock strengths and gas density slopes. Various models, combining shocks with the diffusive shock acceleration and/or adiabatic compression of pre-existing relativistic electrons have already been considered (see e.g. \citealt{Ensslin1998,Roettiger1999,Ensslin2001,Markevitch2005,Kang_Ryu2015}; see also \citealt{Brunetti2014,Weeren2019}). Here we focus on a combination of a pure adiabatic compression (no re-acceleration) with a spherical runaway shock.
We show that such shocks have a characteristic pattern of adiabatic compression/expansion, namely, the region of strong compression is confined to a relatively narrow shell on the downstream side. \cite{Donnert2016} have already considered a set of heuristic models where a parameter characterizing the expansion of the gas on the downstream side was introduced. We argue that this parameter is in fact linked to the Mach number of the shock and we model the evolution of the gas compression factor self-consistently.

This paper is organized as follows. In Section~\ref{sec:analytic}, we present qualitative analytic arguments relevant
to shock propagation in a non-uniform medium. In Section~\ref{sec:numerical}, we show results of numerical simulations, which specifically aim to explore runaway shocks in the context of galaxy clusters and their synchrotron radiation. In Section~\ref{sec:discussion}, we discuss the existence of the ``Habitable zone'' of runaway shocks in the cluster outskirts. In Section~\ref{sec:conclusion}, we summarize our findings.

\section{Shock propagation in a non-uniform medium} \label{sec:analytic}

\begin{figure}
\centering
\includegraphics[width=0.9\linewidth]{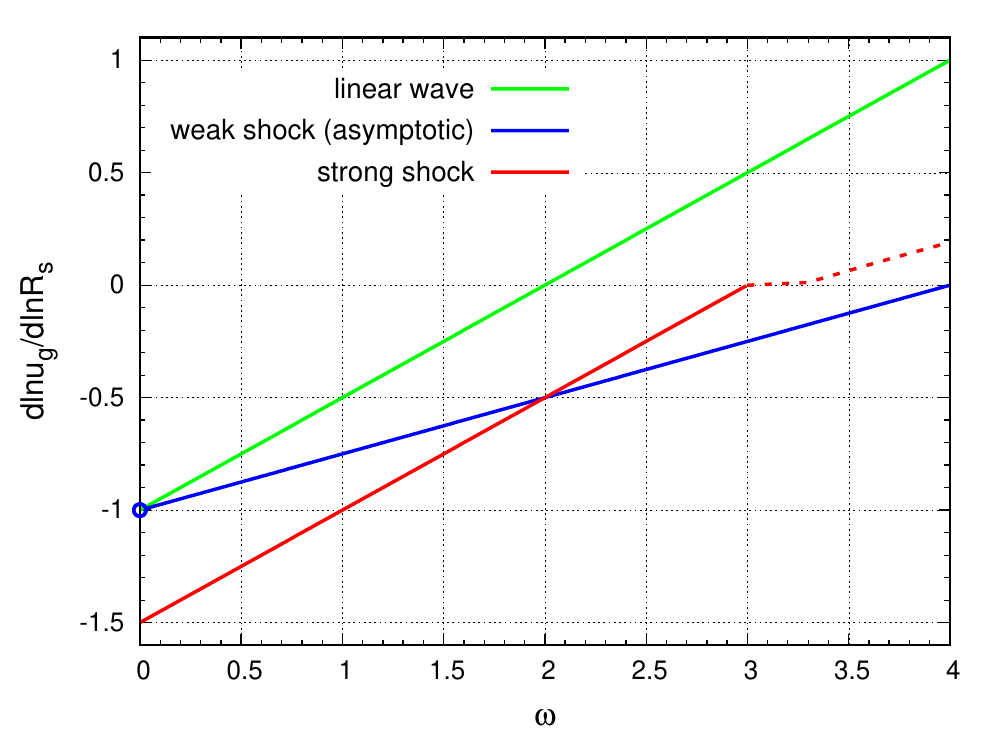}
\caption{Rate of change of gas velocity amplitude  with radius $\eta_{\rm s}=\dd\ln u_{\rm g}/\dd\ln R_{\rm s}$ as a function of the logarithmic slope of the gas density profile $\omega$ (see Eqs.~\ref{eq:law_linear_wave}--\ref{eq:law_strong_shock}). The three curves correspond to linear waves (green line), very weak shocks (blue line), and strong shocks (red lines; the red dashed line shows the results of \citealt{Kushnir2010} for $\omega\geq3$).  Note that, when $\omega=0$, there is no asymptotic power-law solution for weak shocks. This figure shows that all three curves monotonically increase with $\omega$. Both shock curves are lower than that of the linear wave, due to the shock kinetic-energy dissipation (see Section~\ref{sec:analytic}). }
\label{fig:diagram}
\end{figure}

As mentioned above, once the bow shock runs away from its driving subcluster (Fig.~\ref{fig:diagram}), the subsequent shock evolution is  determined by its Mach number and the gas distribution in the atmosphere of the main cluster. During this phase, the motion of the subcluster has essentially no impact on the properties of the shock. Such runaway shocks are reminiscent of the spherical blast waves propagating in a non-uniform medium\footnote{The discussion in this section could also be applied to shocks/sound waves driven by outbursts of supermassive black holes in cluster center, where gas density profiles are shallower.}. It is therefore useful to consider, how a spherical shock is attenuated/amplified when it travels in a  stratified atmosphere. For simplicity, we assume a radial gas density profile of the medium that follows a power law, i.e. $\rho_{\rm gas,0}(r)\propto r^{-\omega}\ (\omega\geq0)$\footnote{The subscript $0$ in gas density $\rho_{\rm gas,0}$ indicates the unperturbed state of the atmosphere.}.

In general, the strength of the propagating spherical shock varies with distance due to three effects: (1) the non-uniform gas density profile, (2) the geometric factor ($\propto r^2$), and (3) the shock dissipation process. For linear waves, only the first two effects are relevant. The energy conservation law (i.e. $\rho_{\rm gas}(R_{\rm s})R_{\rm s}^2u_{\rm g}^2={\rm constant}$) shows that, the velocity amplitude $u_{\rm g}$ of a linear wave changes with radius $R_{\rm s}$ as a power law $u_{\rm g}\propto R_{\rm s}^{\eta_{\rm s}}$, with
\be
\eta_{\rm s} \equiv \frac{\dd\ln u_{\rm g}}{\dd\ln R_{\rm s}} = \frac{\omega}{2} - 1,
\label{eq:law_linear_wave}
\ee
where $u_{\rm g}$ and $c_{\rm s}$ are the gas velocity in the rest frame of the upstream medium and the sound speed, respectively; $R_{\rm s}$ is the position of the wave (see green line in Fig.~\ref{fig:diagram}) or the shock (see discussions below).

Once the wave becomes nonlinear, a discontinuity (weak shock) is formed and dissipation contributes to the decline of the amplitude.
\citet{Landau1945} has shown that for $\omega=0$, the velocity amplitude of a very weak shock, far from its place of origin, declines with distance $R_{\rm s}$ as
\be
u_{\rm g} \propto \frac{1}{R_{\rm s}\sqrt{\ln (R_{\rm s}/\alpha)}},
\ee
where $\alpha$ is a certain constant. Extending  Landau's argument for  $\omega>0$, we find that the asymptotic behavior of a weak shock also follows a power law, whose exponent (see blue line in Fig.~\ref{fig:diagram}) is given as
\be
\eta_{\rm s} = \frac{\omega}{4} - 1.
\label{eq:law_weak_shock}
\ee
It is worth noting that, unlike linear waves, the characteristic length scale of weak shocks changes with distance as  $R_{\rm s}^{\omega/4}$ (in the limit of very large distance). We thus limit our discussion here to  $\omega\leq4$, since for steeper profiles the behavior is more complicated (the shock length scale would become longer than the density scale height of the atmosphere $H_{\rho}\equiv |\dd \ln \rho_{\rm gas,0}/\dd r|^{-1}=r/\omega$ in a finite time).

It is not surprising that strong shocks behave differently from  weak shocks. The former are described by a self-similar (SS) solution if the shock itself is sufficiently strong that the pre-shock pressure can be ignored. In strong shocks, the gas velocity behind the shock is $u_{\rm g}=3V_{\rm s}/4$, where $V_{\rm s}$ is the shock velocity (the adiabatic index $\Gamma=5/3$ is used in this work, unless stated otherwise). When $\omega<3$, a first-type SS solution exists \citep{Sedov1959,Book1994}, which gives
\be
\eta_{\rm s} = \frac{\omega}{2} - \frac{3}{2},
\label{eq:law_strong_shock}
\ee
see red solid line in Fig.~\ref{fig:diagram}. For $\omega<3$, the shock velocity decreases with distance.
When $\omega$ approaches $3$, the shock velocity is expected to remain approximately constant as a function of radius.
When $\omega\geq3$, the situation is different (Eq.~\ref{eq:law_strong_shock} is invalid in this limit), because the gas mass and energy of the atmosphere diverge at the origin ($r=0$). For this case, \citet{Waxman1993} found a second-type SS solution\footnote{More accurately speaking, the second-type SS solution exists when $\omega_{\rm g}<\omega<\omega_{\rm c}$, where $\omega_{\rm g}$ and $\omega_{\rm c}$ are functions of the gas adiabatic index $\Gamma$. For $\Gamma=5/3$, \citet{Waxman1993} showed that $\omega_{\rm g}\simeq3.26$ and $\omega_{\rm c}\simeq7.69$. When $3\leq\omega\leq\omega_{\rm g}$, the change rate $\eta_{\rm s}=0$ gives an asymptotic solution \citep{Kushnir2010}.},
where only the outer part of the flow follows a SS solution and $\eta_{\rm s}$ is found from the numerical solution (e.g., $u_{\rm g}\propto R_{\rm s}^{0.21}$ when $\omega=4$; see red dashed line in Fig.~\ref{fig:diagram}). 

Fig.~\ref{fig:diagram} shows the expected behavior of $\eta_{\rm s}$ as a function of $\omega$ for the three different cases discussed above (namely, a linear wave, a very weak shock, and a strong shock). It is clear that (i) all three curves monotonically increase with $\omega$ and (ii) the values of $\eta_{\rm s}$ for strong and weak shocks are smaller than those for linear waves.

\section{Numerical results} \label{sec:numerical}

The arguments presented in the previous section suggest that a shock, traveling radially, will be attenuated or amplified depending on the strength of the shock and on the slope of the density profile. However,  application of these arguments to merger shocks in clusters faces two difficulties. First, merger shocks usually have moderate Mach numbers (e.g. $\mathcal{M}_{\rm s}\sim2-4$). Neither the results for a very weak shock nor for a strong shock can be directly adopted. Second, weak and strong shocks both require a long period of time (or long travel distance) to evolve towards their asymptotic or SS solutions. This condition, however, might not be satisfied in galaxy clusters. For strong shocks, this issue is particularly serious when $\omega$ is around $3$ (see e.g. \citealt{Kushnir2010}). For these reasons, we perform numerical simulations\footnote{All simulations presented in this section are done by the mesh-based code OpenFOAM (Open Source Field Operation and Manipulation, www.openfoam.org).} in this section to specifically model the motion of runaway shocks in galaxy clusters.

In this study, we applied a 1D blast-wave model, where the shock is initiated by an instantaneous energy release in a small region at the cluster center (see Section~\ref{sec:numerical:blast_wave}).
In all our simulation runs, a static gravitational field with spherical symmetry is assumed to model the cluster environment. Since the radial temperature variations in galaxy clusters (a factor of $\sim2-3$ within the virial radius) are much smaller than the corresponding gas density variations ($2-3$ orders of magnitude, see e.g. \citealt{Ghirardini2019}), the ICM is assumed to be initially isothermal ($T_{\rm gas,0}=1\keV$) and in hydrostatic equilibrium. The corresponding gas density profile follows
\be
\rho_{\rm gas,0}(r)=\rho_{\rm c}\Big(\frac{r}{r_{\rm core}}+1\Big)^{-\omega},
\label{eq:rho_gas}
\ee
where $\rho_{\rm c}=10^7\msun\kpc^{-3}$ and $r_{\rm core}=50\kpc$ are the central density and the core radius of the ICM, respectively. When $r\gg r_{\rm core}$, the logarithmic slope of the gas density profile ($\equiv \dd\ln\rho_{\rm gas,0}(r)/\dd\ln r$) approaches $-\omega$. We note that, the problem discussed here is almost scale-free (radiative cooling is ignored in this study). Our conclusions are only sensitive to the shock Mach number and the gas density slope, but not to the specific choice of parameters (e.g. $T_{\rm gas,0},\ \rho_{\rm c}$) used to describe the cluster atmosphere.

\subsection{1D blast-wave model} \label{sec:numerical:blast_wave}
Our simulation of a 1D blast-wave assumes a spherically symmetric cluster. Four groups of simulations with different initial gas density profiles ($\omega=1,\ 2,\ 3$, and $4$ in Eq.~\ref{eq:rho_gas}, respectively) are performed to study the role of the density profile slope (of the main cluster) on  shock propagation. In each group of runs, blast waves are initiated by a sudden increase of pressure in the innermost cell ($r<1\kpc$) by a factor of $\xi\ (=10^5,\,10^6,\,10^7)$ relative to the equilibrium value. Shocks with moderate Mach numbers $\mathcal{M}_{\rm s}\simeq 1-10$ are formed outside $r_{\rm core}$. The Mach numbers of the excited blast waves when they reach $R_{\rm s}=100$ and $950\kpc$ are summarized in Table~\ref{tab:params}.

\begin{table}
\centering
\begin{minipage}{0.45\textwidth}
\centering
\caption{Parameters of 1D blast-wave simulations.}
\label{tab:params}
\begin{tabular}{*{6}{c}}
  \hline\hline
  IDs &
  $\omega$\footnote{The parameter used in Eq.~(\ref{eq:rho_gas}) to set the slope of the initial gas density profile.} &
  $\xi$\footnote{The factor $\xi$ by which the gas pressure in the innermost cell is increased to initiate the blast wave.} &
  $\mathcal{M}_{\rm s}^{100}$ &
  $\mathcal{M}_{\rm s}^{950}$\footnote{$\mathcal{M}_{\rm s}^{100}$ and $\mathcal{M}_{\rm s}^{950}$ are the Mach numbers of the blast wave when it reaches $R_{\rm s}=100$ and $950\kpc$, respectively. \label{note_ms}} & \\\hline
\hline
  W1P5 & $1$ & $10^5$ & 1.28 & 1.03 \\\hline
  W1P6 & $1$ & $10^6$ & 2.17 & 1.08 \\\hline
  W1P7 & $1$ & $10^7$ & 5.88 & 1.26 \\\hline
  W2P5 & $2$ & $10^5$ & 1.51 & 1.09 \\\hline
  W2P6 & $2$ & $10^6$ & 3.15 & 1.30 \\\hline
  W2P7 & $2$ & $10^7$ & 9.18 & 2.32 \\\hline
  W3P5 & $3$ & $10^5$ & 1.89 & 1.30 \\\hline
  W3P6 & $3$ & $10^6$ & 4.77 & 2.43 \\\hline
  W3P7 & $3$ & $10^7$ & 14.6 & 7.11 \\\hline
  W4P5 & $4$ & $10^5$ & 2.59 & 2.21 \\\hline
  W4P6 & $4$ & $10^6$ & 7.45 & 6.81 \\\hline
  W4P7 & $4$ & $10^7$ & 22.9 & 21.7 \\
\hline\hline
\vspace*{-5mm}
\end{tabular}
\end{minipage}
\end{table}

\begin{figure}
\centering
\includegraphics[width=0.9\linewidth]{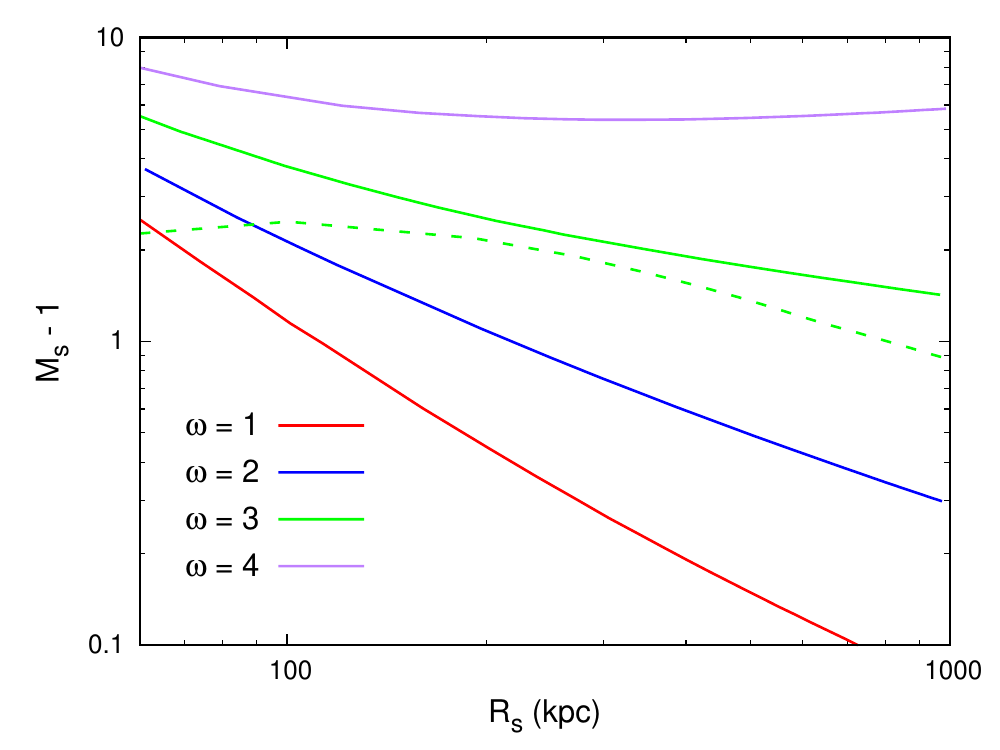}
\caption{Evolution of the Mach number $\mathcal{M}_{\rm s}$ of a shock propagating down the density gradient for different gas density profiles (i.e. different $\omega$, see Eq.~\ref{eq:rho_gas}). $R_{\rm s}$ is the position of the shock front. The solid lines show the evolution of the blast waves from the 1D blast-wave model (runs with $\xi=10^6$; see Section~\ref{sec:numerical:blast_wave}). This figure shows that steep gas density profiles help maintain the strength of  runaway shocks over large distances (green and purple lines).
The green dashed line shows the evolution of the Mach number in simulations of \citealt{Zhang2019}, where the shock is driven by a spherical rigid-body moving through the cluster with $\omega=3$, rather than by a spherical blast wave considered here. The dashed line qualitatively agrees with the result of the blast wave run with the same $\omega$ (green solid line). It shows that, as expected, at large distances from the core, the evolution of the Mach number of the shock does not strongly depend on the way the shock was initiated.}
\label{fig:mach_evolution}
\end{figure}

\begin{figure}
\centering
\includegraphics[width=0.9\linewidth]{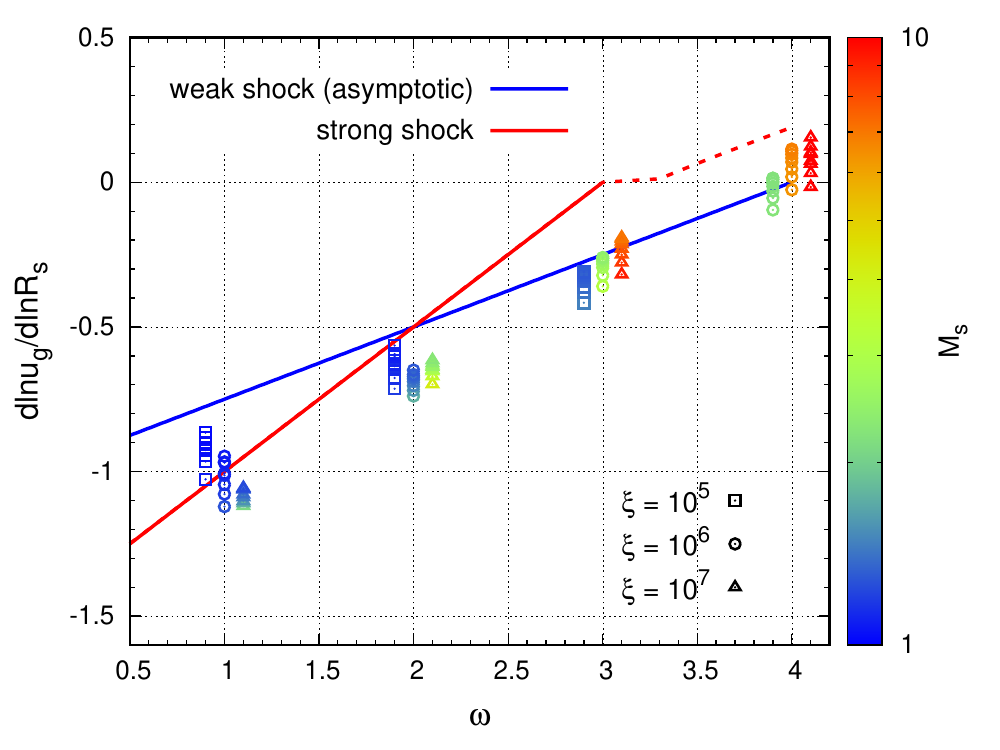}
\caption{Logarithmic derivative of the shock wave amplitude ($\eta_{\rm s}$) in our 1D blast-wave simulations (points). For each of the three values of $\xi$ ($=10^5$, $10^6$, $10^7$) and for each of the four power law slopes $\omega$ of the gas density radial profile used in our simulations, we compute the value of $\eta_{\rm s}$ at nine radii spanning the range from $R_{\rm s} = 200$ to $10^3\kpc$. The colour codes the shock Mach number (the red colour saturates once $\mathcal{M}_{\rm s}>10$; see Table~\ref{tab:params}). For runs with $\xi=10^5/10^7$, the sets of nine points are slightly shifted left/right for display purposes. For comparison, the analytic solutions for the weak and strong shocks are shown with  blue and red lines (these are the same lines as those shown in Fig.~\ref{fig:diagram}). This figure illustrates that numerical simulations are broadly consistent with the theoretical considerations presented in Section~\ref{sec:analytic}, although the match is not perfect (see Section~\ref{sec:numerical:blast_wave}). }
\label{fig:diagram_sim}
\end{figure}

Fig.~\ref{fig:mach_evolution} shows the evolution of the blast wave Mach numbers in runs with $\xi=10^6$ (solid lines). These results follow the trend  discussed in Section~\ref{sec:analytic} (see Fig.~\ref{fig:diagram}), namely, the shallower the gas density profile is, the more rapidly the shock fades with distance. We have further analyzed the rate of the wave amplitude change with radius ($\eta_{\rm s}$ introduced in Section~\ref{sec:analytic}). The values shown in Fig.~\ref{fig:diagram_sim} (symbols) broadly agree with theoretical expectations for the weak and strong shocks (see previous section), which are shown with blue and red lines. In our simulations, all shocks have faster attenuation rates during the initial period than at later times. This is mainly caused by the presence of a core ($r_{\rm core}=50\kpc$) in the adopted gas density profile (see Eq.~\ref{eq:rho_gas}). For example, at $r=100\kpc$, the logarithmic slope of the gas density profile is $\dd\ln\rho_{\rm gas,0}(r)/{\dd\ln r}=-2\omega/3$, i.e. flatter than the asymptotic value ($-\omega$) at large radii. This leads to the faster attenuation of the shock. When the shock leaves  the core ($R_{\rm s}\gg r_{\rm core}$), the velocity attenuation rate approaches theoretical expectations (either for a weak shock or strong shock, depending on the Mach number of the shock). However, we do not expect an exact match between simulations and expectations, like those shown in Fig.~\ref{fig:diagram_sim}. There are a number of reasons for the differences. First, in our simulations the shocks have moderate Mach numbers, which corresponds to neither very strong nor very weak shocks. Secondly, the discussion presented in Section~\ref{sec:analytic} is based on the assumption that the shock has sufficient time to evolve towards the self-similar solution, which is not necessarily achieved in simulations\footnote{The specific condition for the SS solution in our problem is that the shock-swept gas mass is larger than the initial gas mass within the core of the atmosphere. This is also the reason why the numerical solution very slowly approaches the self-similar regime when $\omega\sim3$, as in runs W3P6 and W3P7, shown in Fig.~\ref{fig:diagram_sim}. The shocks still need to propagate a large distance to approach the red line (note that it is just a coincidence that their positions are close to the blue line right now).}. Therefore, detailed simulations of each particular merger configuration are needed, if a quantitative answer is required. However, a qualitative conclusion, that steep density profiles help to prevent attenuation of shocks, of course, holds. Accordingly,  runaway shocks are expected to be ``long-lived'' outside $R_{\rm 500}$ where the logarithmic slope of the gas density profile of the atmosphere (i.e. $-\omega$) is steeper than $\sim-3$. The above discussion assumes isothermal gas over the entire radial range, so that the velocity and the Mach number are unambiguously related. In real clusters, the temperature declines beyond $\sim R_{500}$ \citep[e.g.][]{Vikhlinin2005}. This leads to an additional increase of the Mach number for  shocks propagating to the outskirts of clusters \citep[e.g.][]{Ha2018}.

\begin{figure}
\centering
\includegraphics[width=0.9\linewidth]{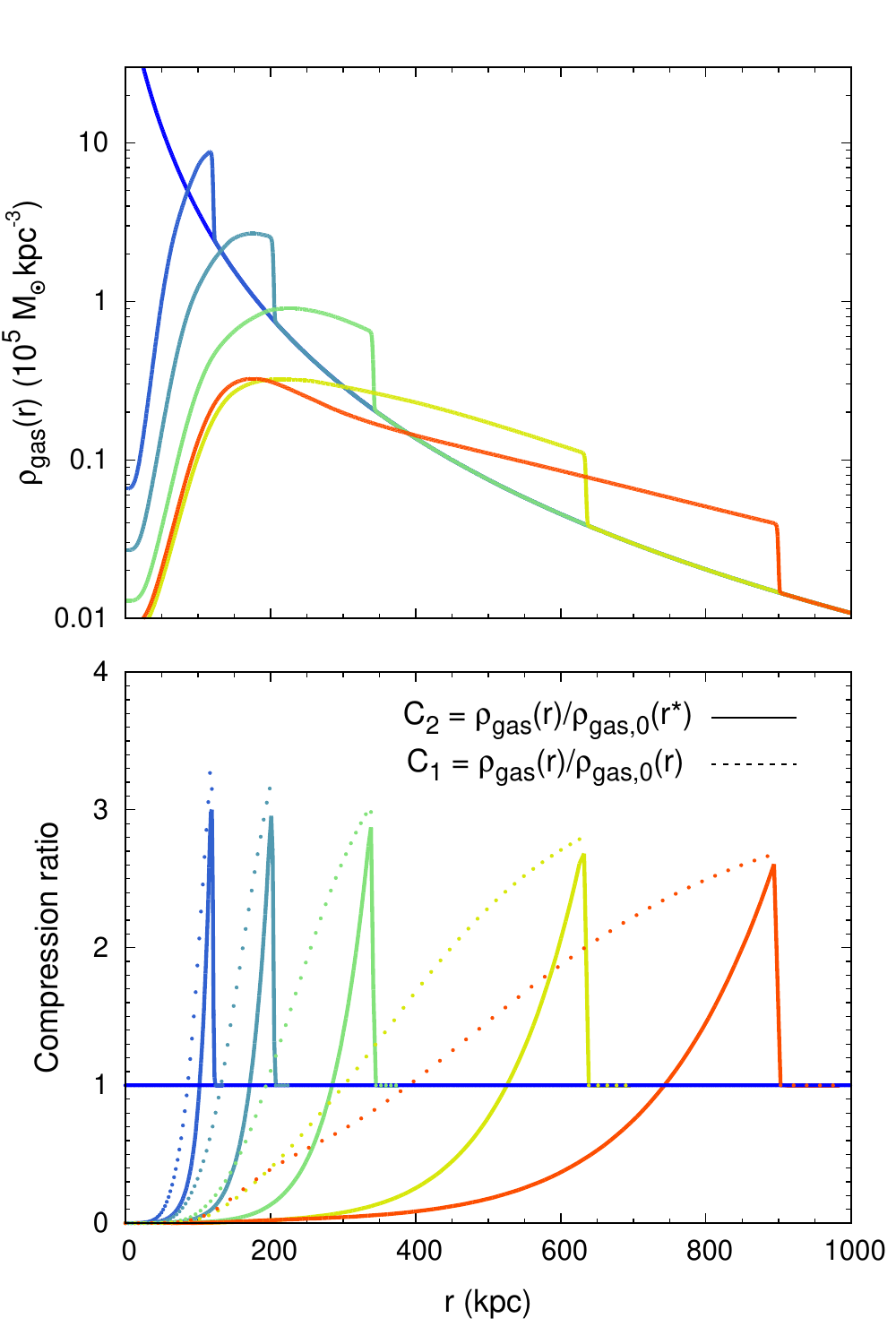}
\caption{\textit{Top panel:} evolution of the gas density profile in the run W3P6 ($\omega=3$). During late evolutionary phases, the density profiles, inside and outside the shock front, resemble power laws, albeit with slightly different indices.  \textit{Bottom panel:} the ``Lagrangian'' gas compression ratio $C_2=\rho_{\rm gas}(r)/\rho_{\rm gas,0}(r^\ast)$ (solid lines; see Eq.~\ref{eq:integral_eq} for the definition of $r^\ast$). For comparison, the ``Eulerian'' compression ratio $C_1=\rho_{\rm gas}(r)/\rho_{\rm gas,0}(r)$ is shown with the dotted lines. The $C_2$ profiles appear much  narrower than the $C_1$ profiles (see Section~\ref{sec:numerical:blast_wave}). Note that due to a minor numerical artifact, $C_1$ appears slightly higher than $C_2$ at the shock front, while by design $C_2\equiv C_1$ at this point.}
\label{fig:shock_profile}
\end{figure}

\begin{figure}
\centering
\includegraphics[width=0.9\linewidth]{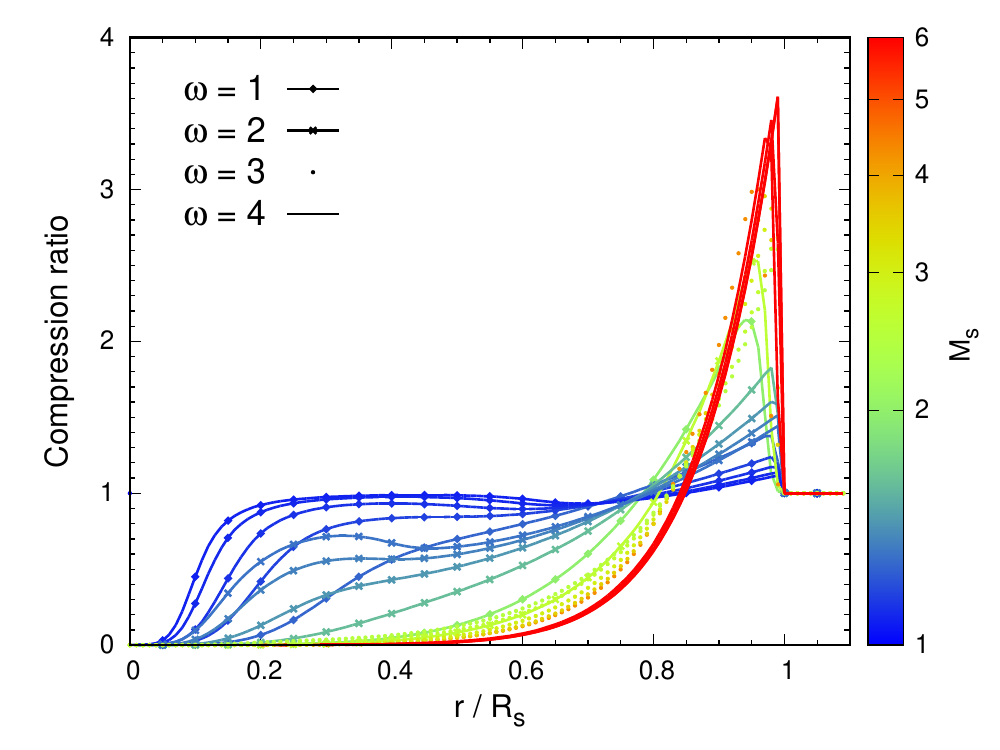}
\caption{Gas compression ratio (factor $C_2$) profiles of the 1D blast waves propagating in gas density profiles with different slopes. The magnitude of the initial energy release is set by the same value of $\xi=10^6$ (for each run, a series of curves at different time are shown; see text for details). The colour codes the shock Mach number. Once the Mach number is larger than $\sim 3$, the compression ratio profile depends primarily on the Mach number rather than on the slope of the density profile (see Section~\ref{sec:numerical:blast_wave}). }
\label{fig:comp_ratio}
\end{figure}

Finally, we consider the shock structure in the blast-wave model, which is important for interpretation of the synchrotron emission (i.e. radio relics) associated with the merger shocks in galaxy clusters (see also Section~\ref{sec:numerical:radio_emission}). The top panel in Fig.~\ref{fig:shock_profile} shows the evolution of the gas density profile in the run W3P6 with $\omega=3$. The profiles corresponding to later phases (see e.g. the red curve) are of particular interest here. As expected, the shock strength only weakly declines with radius, the gas profile has a characteristic jump at the shock and a smooth density distribution before and after the shock. The density jump, at the shock, depends on the adiabatic index and is, of course, a function of the shock Mach number, which can be derived from X-ray data. The question arises if the same density jump (or, equivalently, the compression factor) applies to the entire downstream region. Two versions of the compression factors are explored in this work (shown in the bottom panel in Fig.~\ref{fig:shock_profile}).
\begin{itemize}
  \item The first version, that we call $C_1$, is the ratio of the gas density at a given radius to the initial density at the same radius, i.e.,  $C_1=\rho_{\rm gas}(r)/\rho_{\rm gas,0}(r)$. This ratio (shown by the dotted lines in Fig.~\ref{fig:shock_profile}) characterizes the modification of the density profile by the runaway shock. Since $C_1$ slowly decreases towards the cluster center, the density profile seen in the top panel appears slightly shallower than the initial density profiles. This justifies the use, by X-ray astronomers, of an analytic model consisting of two power laws with slightly different slopes for fitting the shocks \citep{Markevitch2007}.
  \item The second version of the compression factor, that we call $C_2$, characterizes the compression in the Lagrangian sense, namely the ratio of the gas density to the initial gas density {\it of the same gas lump}, i.e.,  $C_2=\rho_{\rm gas}(r)/\rho_{\rm gas,0}(r^\ast)$, where $r^\ast$ can be found from the following integral equation:
\be
\int_{0}^{r}{\rho_{\rm gas}(r')r'^2\dd r'} = \int_{0}^{r^\ast}{\rho_{\rm gas,0}(r')r'^2\dd r'}.
\label{eq:integral_eq}
\ee
The $C_2$ profiles in the run W3P6 are shown in the bottom panel of Fig.~\ref{fig:shock_profile} (solid lines). While $C_2\equiv C_1$ at the shock front, the width of region with large $C_2$ is clearly much smaller than that for $C_1$.
Therefore, relative to the initial state of gas, the strong compression is in a narrow shell with width $\sim R_{\rm s}/5$ just behind the shock. At larger distances from the shock, the gas density of a given lump is even smaller than it was before being swept by the shock. Fig.~\ref{fig:comp_ratio} shows that this result is not sensitive to $\omega$ rather, but depends on the shock Mach number (saturates for $\mathcal{M}_{\rm s}>3$) and reflects the spherical geometry of the problem.
Indeed, unlike the case of a plane-parallel shock, in spherical geometry, the gas on the downstream side is expected to expand. \citet{Donnert2016} modelled the structure of the shock in the CIZA~J2242.8+5301 cluster with an ad hoc model, that uses a parameter $t_{\rm exp}$ to account for adiabatic expansion (see their section~2.1). In their model, $t_{\rm exp}$ is essentially a free parameter. Figs.~\ref{fig:shock_profile} and \ref{fig:comp_ratio} show that, in fact, the compression ratio is tightly linked to the Mach number, removing the uncertainty associated with the choice of $t_{\rm exp}$. Moreover, the compression ratio drops below unity for $M\gtrsim 2$ at $r\sim 0.8R_{\rm s}$, when the expansion is accounted for self-consistently. An approximate comparison with the \citet{Donnert2016} model shows that the size of the compressed region is a factor of $2-3$ smaller than the value expected for  $t_{\rm exp}=240\Myr$ used for the CIZA~J2242.8+5301 shock.
\end{itemize}

\subsection{Synchrotron radiation} \label{sec:numerical:radio_emission}
Even though our simulations are purely hydrodynamic, we can assess the expected impact of propagating shocks on the synchrotron emission under the assumption that relativistic particles and magnetic fields have no direct impact on the dynamics of the gas.

Since the shocks considered here have modest Mach numbers $\lesssim2-3$, the particle acceleration efficiency might be low \citep[e.g.][]{Kang2015}. We therefore assume that no acceleration takes place and focus exclusively on the adiabatic compression of pre-existing (fossil) relativistic electrons as the mechanism to boost the synchrotron emissivity in the downstream region \citep[see e.g.][]{Laan1962,Ensslin2001,Ensslin2002,Pfrommer2011}. Furthermore, we assume that these fossil electrons are uniformly distributed in the unshocked gas, and initially have a power-law energy distribution, i.e. ${\rm d}n_{e}/{\rm d}\gamma\propto\gamma^{-p}$, where $p=3.5$ and $\gamma$ is the particle Lorentz factor.  The relativistic electrons are assumed to be isotropic in the upstream and downstream regions.

We assume that the magnetic field, on the upstream side of the shock, is fixed at $B_0=2{\,\rm\mu G}$. For the downstream side, the magnetic field strength is set to $B(r)\propto B_0 C_2(r)^{2/3}$, which corresponds to the case of relativistic gas with adiabatic index $\Gamma=4/3$ with compression ratio $C_2$.

While the above assumptions are clearly idealized, they serve the purpose of illustrating the magnitude of the synchrotron emissivity boost factor $W_\nu$ at frequency $\nu$ for a given gas lump, after it passes through the shock. The time evolution of the synchrotron emission subject to the adiabatic compression and inverse Compton (IC) and synchrotron energy losses, can be easily calculated, if the time evolution of the compression factor is known, which in our case is given by function $C_2(t)$ \citep[e.g.][]{Matthews1990,Churazov2001,Ensslin2001}. This allows us to recover the time evolution of the boost factor $W_\nu$.

\begin{figure}
\centering
\includegraphics[width=0.9\linewidth]{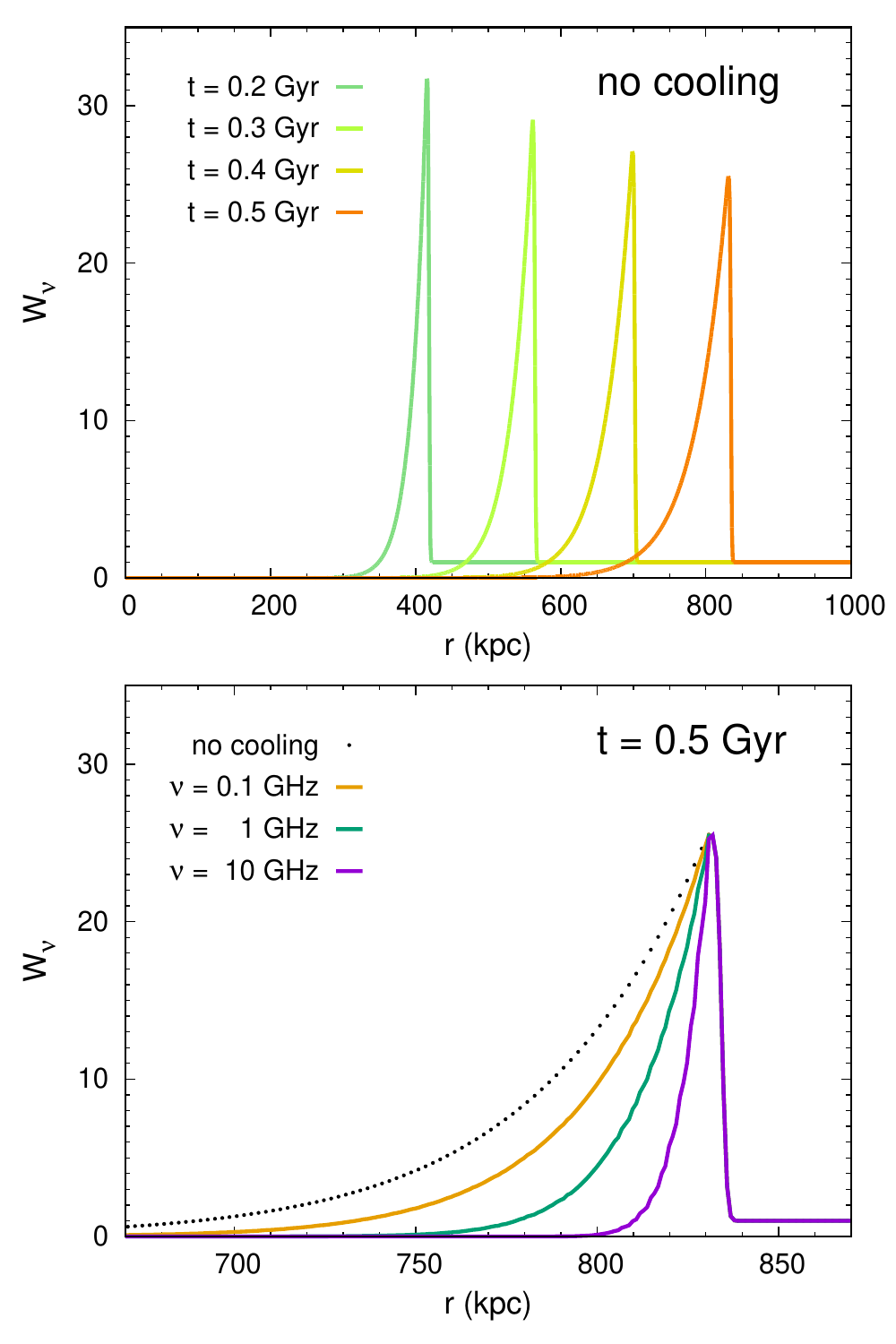}
\caption{Synchrotron emissivity boost factor $W_\nu$ evaluated for the run W3P6 ($\omega=3$). By design, $W_\nu\equiv 1$ in the upstream region. \textit{Top panel:} time evolution of $W_\nu$, when only adiabatic compression is considered (no synchrotron or IC losses). In this case, $W_\nu$ is set by the instantaneous value of the gas compression factor $C_2(r)$ (see bottom panel of Fig.~\ref{fig:shock_profile}). \textit{Bottom panel:} $W_\nu$ at different emitted frequencies ($t=0.5\Gyr$), when the cooling losses are included (solid lines). As expected, the profiles are narrower at a higher frequencies. This figure shows that the synchrotron emission due to adiabatic compression of pre-existing relativistic electrons by the runaway merger shocks is confined to a narrow radial shell (see Section~\ref{sec:numerical:radio_emission}). }
\label{fig:radio_emission}
\end{figure}

Fig.~\ref{fig:radio_emission} shows the synchrotron emissivity boost factor profiles for the simulation run W3P6. By design, the scale factor is equal to unity in the upstream region. The top panel shows the time evolution of the $W_\nu$ profiles, when the cooling processes (synchrotron radiation and IC scattering) are not included. In the absence of cooling losses, the adiabatic compression preserves the shape of the radiation spectrum but only modifies its amplitude by a frequency-independent factor $W_{\nu}=C_2(r)^{1+2p/3}$ \citep{Markevitch2005}. This plot shows that, even without cooling losses, the strongly enhanced synchrotron emission is confined to a relatively narrow shell of compressed gas and drops in the downstream region due to gas re-expansion. In projection, such structures might appear as a bright rim with a sharper outer edge and somewhat smoother inner one.

If the effects of cooling losses are included, the profiles become even narrower. This is illustrated in the bottom panel of Fig.~\ref{fig:radio_emission}. For simplicity, we turn on the cooling losses only on the downstream side of the shock. Since higher energy electrons have shorter lifetimes and the characteristic frequency of the synchrotron emission scales as $\propto\gamma^2$, the width of the enhanced emissivity region decreases with the emitted frequency $\nu$.

Thus, the structure of the runaway merger shocks can naturally produce peaked emissivity profiles even when the cooling losses can be neglected. This naturally follows from a sequence of compression at the shock and successive expansion of the gas. The above discussion is of course a qualitative one. We defer the discussion of real clusters, the analysis of the magnetic field topology, and projection effect to our forthcoming paper.

\section{``Habitable zone'' of runaway merger shocks} \label{sec:discussion}

In Sections~\ref{sec:analytic} and \ref{sec:numerical}, we have shown that the fate of  runaway merger shocks, traveling through the ICM, is sensitive to the steepness of the gas density profile of the atmosphere. These shocks could be long-lived if the gas density profile is as steep as $\sim r^{-2}-r^{-4}$. It is therefore of considerable interest to discuss the possible ``Habitable zone'' of runaway merger shocks existing in galaxy clusters.

\begin{figure}
\centering
\includegraphics[width=0.9\linewidth]{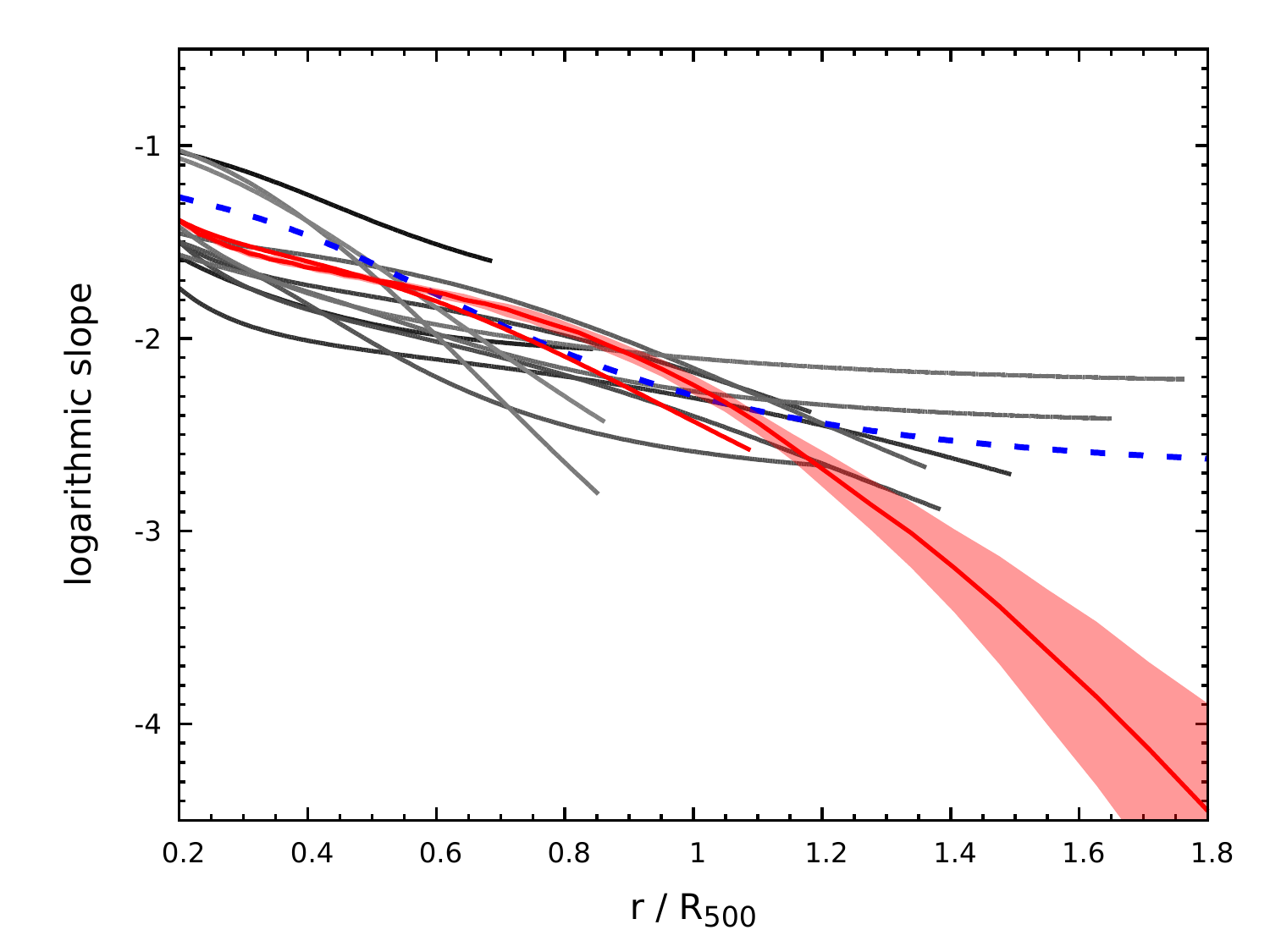}
\caption{Logarithmic slope of the gas density profile ($\equiv\dd\ln\rho_{\rm gas}(r)/\dd\ln r$) as a function of cluster radius for 12 galaxy clusters (solid lines, presented in \citealt{Vikhlinin2006}). These curves are truncated at the detection radius $r_{\rm det}$ where X-ray brightness is detected at $>3\sigma$, or the outer boundary of the \textit{Chandra} field of view for the corresponding cluster. The results for the cluster A133 from \citet{Vikhlinin2006} is highlighted in red, and is compared with that from a deeper \textit{Chandra} observation with an exposure time $\rm\simeq2\,Ms$ (red curve with a $1\sigma$ error band, A.~Vikhlinin, priv. comm.). The result from the deeper observation shows a steeper gas density profile, e.g. $\sim r^{-4}$ at $1.6R_{500}$ (all resolved bright clumps and filaments are excised). As a comparison, the slope of the X-COP sample mean profile is shown as the blue dashed line \citep{Ghirardini2019}. This figure shows that the gas density profile of galaxy clusters is generally steeper than $\sim r^{-2.5}$ at $r\gtrsim R_{500}$, where a ``Habitable zone'' of runaway merger shocks might exist (see Section~\ref{sec:discussion}).}
\label{fig:slope}
\end{figure}

Fig.~\ref{fig:slope} shows the logarithmic slope of the gas density profile as a function of radius for 12 galaxy clusters (solid lines, presented in \citealt{Vikhlinin2006}), derived from the cluster best-fit model (eq.~3 in \citealt{Vikhlinin2006}, but the second (inner) $\beta$-model component is not included, because we focus here on the outer regions). $Chandra$ and $ROSAT$ data were used in the fitting. These curves are truncated at the detection radius $r_{\rm det}$ (see table~1 in \citealt{Vikhlinin2006}), far beyond which the extrapolation becomes unreliable. We can see that, within $0.5R_{500}$, the gas density profiles are generally shallower than $r^{-2}$. If the shock becomes detached from the subcluster in this region, then its amplitude is expected to decrease with radius as $\propto r^{-1/2}$.
When $r$ approaches $R_{500}$, the profile becomes progressively steeper, with the logarithmic slope between $-2$ and $-3$. Beyond $R_{500}$, the slope continues to steepen for most of the cluster sample \citep[see also][]{Morandi2015}. We stress here that cool, dense clumps, and filaments complicate accurate measurements of the X-ray profiles  \citep{Nagai2011,Simionescu2011,Zhuravleva2013,Avestruz2014} and might lead to a shallower profile than that of the dominant (by volume) component, which has lower gas density. This problem can be overcome with sensitive, high-angular resolution observations. To this end, we show the results of  deeper \textit{Chandra} observation of A133 with total exposure time $\rm \approx 2\,Ms$ (red curve with an error band, A.~Vikhlinin, priv. comm.; see also \citealt{Morandi2014}), which suggest that the slope steepens to values between $-3$ and $-4$, once all resolved bright clumps and filaments are excised. Cosmological simulations also suggest very steep slopes of the cluster gas density profiles ($\sim -3$ or steeper) around the virial radius \citep[see e.g.][]{Roncarelli2006,Vazza2010,Lau2015}. The profiles might be even steeper around the cluster splashback radius (as steep as $-4$; see e.g. \citealt{Diemer2014,Lau2015}).

All these results suggest that a ``Habitable zone'' populated by moderately strong shocks could exist outside $\simeq R_{500}$, where runaway merger shocks do not attenuate quickly. While these shocks might be difficult to detect in X-rays, they might be responsible for powering radio relics often found in the periphery of merging clusters (see discussions in Section~\ref{sec:numerical:radio_emission}). In this regard, radio relics, discovered in cluster peripheries (e.g. A746 \citealt{Weeren2011}; A2744 \citealt{Pearce2017}; see also \citealt{Weeren2019} for a review), provide a unique opportunity to locate and study runaway merger shocks, whose X-ray signals are too weak to be detected. Studying such shocks is important to our understanding the merger configurations of galaxy clusters. For example, the southwestern radio relics, discovered in the Coma cluster \citep{Brown2011,Akamatsu2013,Erler2015}, support the post-merger scenario proposed by \citet{Lyskova2019}, where a runaway shock (consistent with the position of observed radio relics) is generated by the infalling galaxy group associated with NGC~4839.

Finally, we suggest that the differences in the gas profile slopes  might provide a natural explanation for the ``segmented'' appearance of merger shocks discussed in \citet{Paul2011}. We know that, not very far from the virial radius, the matter distributions in galaxy clusters are highly inhomogeneous/asymmetric. The gas density profile is shallower along the direction of filaments than between them \citep[e.g. see][]{Sato2012}. Along the filaments, runaway merger shocks would propagate more slowly and attenuate more quickly, while between the filaments, the shocks would move more quickly and weaken more slowly. As a consequence, these runaway shocks would visually break into separate segments/wedges and appear more prominent in the ``valleys'' between the high-density filaments. The ``surviving'' segmented shocks might further interact with virial shocks and form more complicated structures in cluster outskirts (e.g. cold fronts, see \citealt{Birnboim2010}; see also e.g. \citealt{Paul2011,Ha2018} for discussions in the context of cosmological simulations).

\section{Conclusion} \label{sec:conclusion}

In this paper, we have explored the propagation of runaway merger shocks in galaxy clusters. Our findings can be summarized as follows:
\begin{itemize}
\item The evolution of merger shocks can be described as a two-phase process: the ``driven phase'' and the ``detached phase''. In the detached phase, the subcluster, which drives a bow shock, is decelerated by the gravity of the main cluster, but the shock continues on its way towards the main-cluster outskirts. Meanwhile, the steep gas density profiles help the runaway shocks maintain their strength or even amplify them (if the profile is as steep as, or steeper than, $\sim r^{-3}$).
\item High angular-resolution X-ray observations show that, beyond $R_{500}$, the gas density profile of galaxy clusters could indeed be as steep as $\sim r^{-3}$. This suggests that a ``Habitable zone'' could exist in the cluster outskirts, where runaway merger shocks are relatively ``long-lived''. Since the gas density in this region is very low, the detection of these shocks in X-ray is challenging.
\item A characteristic feature of runaway shocks with $\mathcal{M}_{\rm s}\gtrsim2$ is that the strongly compressed gas, relative to its initial state, is confined to a relatively narrow region just behind the shock (similar to the case of a spherical shock moving in a uniform medium). Namely, a strong compression at the shock front is followed by a strong decompression. If adiabatic compression plays a role in boosting the radio emissivity of radio relics, then even in the absence of synchrotron and IC losses, the radio emission is expected to be confined to this shell.
\end{itemize}

\section*{Acknowledgements}

We are grateful to Alexey Vikhlinin for providing us the A133 density profile. CZ thanks Daisuke Nagai for helpful discussions and comments on the draft. EC and NL acknowledge support from the Russian Science Foundation grant 19-12-00369. WF acknowledges support from the Smithsonian Institution, NASA Contract NAS-08060, and NASA Grant GO2-13005X.





\bsp	
\label{lastpage}

\begin{thebibliography}{99}

\bibitem[Akamatsu et al.(2013)]{Akamatsu2013} Akamatsu, H., Inoue, S., Sato, T., et al.\ 2013, Publications of the Astronomical Society of Japan, 65, 89.

\bibitem[Avestruz et al.(2014)]{Avestruz2014} Avestruz, C., Lau, E.~T., Nagai, D., \& Vikhlinin, A.\ 2014, \apj, 791, 117

\bibitem[Bautz et al.(2009)]{Bautz2009} Bautz, M.~W., Miller, E.~D., Sanders, J.~S., et al.\ 2009, Publications of the Astronomical Society of Japan, 61, 1117.

\bibitem[Birnboim et al.(2010)]{Birnboim2010} Birnboim, Y., Keshet, U., \& Hernquist, L.\ 2010, \mnras, 408, 199.

\bibitem[Book(1994)]{Book1994} Book, D.~L.\ 1994, Shock Waves, 4, 1

\bibitem[Borgani \& Kravtsov(2011)]{Borgani2011} Borgani, S., \& Kravtsov, A.\ 2011, Advanced Science Letters, 4, 204

\bibitem[Brown \& Rudnick(2011)]{Brown2011} Brown, S., \& Rudnick, L.\ 2011, \mnras, 412, 2.

\bibitem[Brunetti, \& Jones(2014)]{Brunetti2014} Brunetti, G., \& Jones, T.~W.\ 2014, International Journal of Modern Physics D, 23, 1430007-98

\bibitem[Bykov et al.(2015)]{Bykov2015} Bykov, A.~M., Churazov, E.~M., Ferrari, C., et al.\ 2015, \ssr, 188, 141

\bibitem[Churazov et al.(2001)]{Churazov2001} Churazov, E., Br{\"u}ggen, M., Kaiser, C.~R., et al.\ 2001, \apj, 554, 261

\bibitem[Diemer \& Kravtsov(2014)]{Diemer2014} Diemer, B., \& Kravtsov, A.~V.\ 2014, \apj, 789, 1

\bibitem[Donnert et al.(2016)]{Donnert2016} Donnert, J.~M.~F., Stroe, A., Brunetti, G., Hoang, D., \& Roettgering, H.\ 2016, \mnras, 462, 2014

\bibitem[En{\ss}lin et al.(1998)]{Ensslin1998} Ensslin, T.~A., Biermann, P.~L., Klein, U., et al.\ 1998, \aap, 332, 395

\bibitem[En{\ss}lin \& Gopal-Krishna(2001)]{Ensslin2001} En{\ss}lin, T.~A., \& Gopal-Krishna 2001, \aap, 366, 26

\bibitem[En{\ss}lin \& Br{\"u}ggen(2002)]{Ensslin2002} En{\ss}lin, T.~A., \& Br{\"u}ggen, M.\ 2002, \mnras, 331, 1011.

\bibitem[Erler et al.(2015)]{Erler2015} Erler, J., Basu, K., Trasatti, M., Klein, U., \& Bertoldi, F.\ 2015, \mnras, 447, 2497

\bibitem[Feretti et al.(2012)]{Feretti2012} Feretti, L., Giovannini, G., Govoni, F., \& Murgia, M.\ 2012, \aapr, 20, 54

\bibitem[George et al.(2009)]{George2009} George, M.~R., Fabian, A.~C., Sanders, J.~S., et al.\ 2009, \mnras, 395, 657.

\bibitem[Ghirardini et al.(2019)]{Ghirardini2019} Ghirardini, V., Eckert, D., Ettori, S., et al.\ 2019, \aap, 621, A41.

\bibitem[Ha et al.(2018)]{Ha2018} Ha, J.-H., Ryu, D., \& Kang, H.\ 2018, \apj, 857, 26

\bibitem[Hong et al.(2014)]{Hong2014} Hong, S.~E., Ryu, D., Kang, H., et al.\ 2014, \apj, 785, 133.

\bibitem[Kang(2015)]{Kang2015} Kang, H.\ 2015, Journal of Korean Astronomical Society, 48, 155.

\bibitem[\protect\citeauthoryear{Kang \& Ryu}{2015}]{Kang_Ryu2015} Kang H., Ryu D., 2015, ApJ, 809, 186

\bibitem[Kushnir \& Waxman(2010)]{Kushnir2010} Kushnir, D., \& Waxman, E.\ 2010, \apj, 723, 10

\bibitem[Landau(1945)]{Landau1945} Landau, L.~D.\ 1945,  J. Phys. USSR, 9, 496

\bibitem[Lau et al.(2015)]{Lau2015} Lau, E.~T., Nagai, D., Avestruz, C., Nelson, K., \& Vikhlinin, A.\ 2015, \apj, 806, 68

\bibitem[Lyskova et al.(2019)]{Lyskova2019} Lyskova N., Churazov E., Zhang C., Forman W., Jones C., Dolag K., Roediger E., Sheardown A., 2019, MNRAS, 485, 2922

\bibitem[Markevitch et al.(2005)]{Markevitch2005} Markevitch, M., Govoni, F., Brunetti, G., et al.\ 2005, \apj, 627, 733

\bibitem[Markevitch \& Vikhlinin(2007)]{Markevitch2007} Markevitch M., Vikhlinin A., 2007, \physrep, 443, 1

\bibitem[Matthews \& Scheuer(1990)]{Matthews1990} Matthews A.~P., Scheuer P.~A.~G., 1990, MNRAS, 242, 616

\bibitem[Morandi \& Cui(2014)]{Morandi2014} Morandi, A., \& Cui, W.\ 2014, \mnras, 437, 1909

\bibitem[Morandi et al.(2015)]{Morandi2015} Morandi, A., Sun, M., Forman, W., \& Jones, C.\ 2015, \mnras, 450, 2261

\bibitem[Nagai \& Lau(2011)]{Nagai2011} Nagai, D., \& Lau, E.~T.\ 2011, \apjl, 731, L10

\bibitem[Okabe et al.(2014)]{Okabe2014} Okabe, N., Umetsu, K., Tamura, T., et al.\ 2014, \pasj, 66, 99

\bibitem[Paul et al.(2011)]{Paul2011} Paul, S., Iapichino, L., Miniati, F., Bagchi, J., \& Mannheim, K.\ 2011, \apj, 726, 17

\bibitem[Pearce et al.(2017)]{Pearce2017} Pearce, C.~J.~J., van Weeren, R.~J., Andrade-Santos, F., et al.\ 2017, \apj, 845, 81

\bibitem[Pfrommer \& Jones(2011)]{Pfrommer2011} Pfrommer, C., \& Jones, T.~W.\ 2011, \apj, 730, 22.

\bibitem[Ricker \& Sarazin(2001)]{Ricker2001} Ricker, P.~M., \& Sarazin, C.~L.\ 2001, \apj, 561, 621

\bibitem[Roettiger et al.(1999)]{Roettiger1999} Roettiger, K., Burns, J.~O., \& Stone, J.~M.\ 1999, \apj, 518, 603

\bibitem[Roncarelli et al.(2006)]{Roncarelli2006} Roncarelli, M., Ettori, S., Dolag, K., et al.\ 2006, \mnras, 373, 1339

\bibitem[Sato et al.(2012)]{Sato2012} Sato, T., Sasaki, T., Matsushita, K., et al.\ 2012, \pasj, 64, 95

\bibitem[Sedov(1959)]{Sedov1959} Sedov, L.~I.\ 1959, Similarity and Dimensional Methods in Mechanics, New York: Academic Press, 1959

\bibitem[Simionescu et al.(2011)]{Simionescu2011} Simionescu, A., Allen, S.~W., Mantz, A., et al.\ 2011, Science, 331, 1576

\bibitem[van der Laan(1962)]{Laan1962} van der Laan, H.\ 1962, \mnras, 124, 179

\bibitem[van Weeren et al.(2010)]{Weeren2010} van Weeren, R.~J., R{\"o}ttgering, H.~J.~A., Br{\"u}ggen, M., \& Hoeft, M.\ 2010, Science, 330, 347

\bibitem[van Weeren et al.(2011)]{Weeren2011} van Weeren, R.~J., Br{\"u}ggen, M., R{\"o}ttgering, H.~J.~A., et al.\ 2011, \aap, 533, A35

\bibitem[van Weeren et al.(2019)]{Weeren2019} van Weeren, R.~J., de Gasperin, F., Akamatsu, H., et al.\ 2019, \ssr, 215, 16

\bibitem[Vazza et al.(2009)]{Vazza2009} Vazza, F., Brunetti, G., \& Gheller, C.\ 2009, \mnras, 395, 1333

\bibitem[Vazza et al.(2010)]{Vazza2010} Vazza, F., Brunetti, G., Gheller, C., \& Brunino, R.\ 2010, \na, 15, 695

\bibitem[Vazza et al.(2012)]{Vazza2012} Vazza, F., Br{\"u}ggen, M., van Weeren, R., et al.\ 2012, \mnras, 421, 1868

\bibitem[Vikhlinin et al.(2005)]{Vikhlinin2005} Vikhlinin A., Markevitch M., Murray S.~S., Jones C., Forman W., Van Speybroeck L., 2005, ApJ, 628, 655

\bibitem[Vikhlinin et al.(2006)]{Vikhlinin2006} Vikhlinin, A., Kravtsov, A., Forman, W., et al.\ 2006, \apj, 640, 691

\bibitem[Waxman \& Shvarts(1993)]{Waxman1993} Waxman, E., \& Shvarts, D.\ 1993, Physics of Fluids A, 5, 1035

\bibitem[Zhang et al.(2014)]{Zhang2014} Zhang, C., Yu, Q., \& Lu, Y.\ 2014, \apj, 796, 138.

\bibitem[Zhang et al.(2019)]{Zhang2019} Zhang, C., Churazov, E., Forman, W.~R., \& Jones, C.\ 2019, \mnras, 482, 20

\bibitem[Zhuravleva et al.(2013)]{Zhuravleva2013} Zhuravleva, I., Churazov, E., Kravtsov, A., et al.\ 2013, \mnras, 428, 3274
\end{thebibliography}
\end{document}